# Communication on the letter: "Evolutionary Conservation of Motif Constituents in the Yeast Protein Interaction Network."[1]

Stan Bumble, Dept. of Physics, Community College of Philadelphia,
Philadelphia, PA 19130, sbumble@ccp.edu

**Abstract:**
Former work on an application of order-disorder theory is recalled as a vehicle to add further development and significance to the recent paper on motifs in protein interactions.

In the above notable and perspicacious letter the authors indicate that motifs may represent evolutionary conserved topological units of cellular networks formed with specific participatory biological function characteristics. I call their attention to a paper[2] and thesis[3]. There a treatment of order-disorder theory is applied to a study of gas adsorption phenomena on a triangular lattice. The treatment follows that of Hijmans and de Boer[4], which is well delineated. The probability of various occupation states of progressive geometric figures (point, bond, triangle, rhombus) are derived as approximations using rigorous statistical mechanics and thermodynamics for this application, as a function of interaction energy, coordination number and pressure (chemical potential). Such a model was later applied[5,6,7] to genetic or protein networks. It can enrich your findings in the evolutionary conservation of motifs in protein interaction networks for organisms including *Saccharomyces Cerevisiae* and humans. Although the motifs in the letter include the bond, triangle, square and pentagon, the method described[2] can be extended. Results may lead to values for the interaction energy and coordination numbers of the protein networks. There is also, in reference 3, pathway selection information in the equilibrium case for occupied states of the rhombus figure, which can have numerical significance for the motif analogy. There are also entropy considerations, mentioned there, when motifs are selected from the rest of the network, that should not be ignored.

The gas adsorption application has been well shown to be valid for the bond approximation by Bethe's work[8] as well as by ample experimentation. It can be visualized as a "virtual" detailed network between two phases: one a gas phase; the other a condensed phase, that brings about the occupation of states of the condensed phase as controlled by the geometry of the unoccupied lattice, the gas pressure (or chemical potential) and the nature of the gas molecules.

This real occupational lattice network suggests a connection with the virtual global scenario for the cell and its genetic model with the original DNA-genetic source and the phenotype end products and the proteins in the network acting as "movers". Is it possible that a certain confluence of chemical potential (pressure), interaction energy and lattice structure in the gas adsorption application that leads to critical phenomena, (i.e., the condensed adsorbed molecules separate into two phases) has an analogy in the biological case? If so, what are the evolutionary phases in the biological application?